\documentclass[12pt,authoryear]{elsarticle}

\usepackage{amssymb}
\usepackage{amsthm}
\usepackage{amsmath}
\usepackage{mathrsfs}
\usepackage{graphicx}
\usepackage{epstopdf}
\usepackage{float}
\usepackage{caption}
\usepackage{subcaption}
\usepackage{bm}
\usepackage{bbm}
\usepackage{mathrsfs}
\usepackage{cleveref}
\usepackage{color,soul} 
\usepackage{color} 
\usepackage{accents}
\usepackage[export]{adjustbox} 
\soulregister\citep7 
\soulregister\citet7 
\soulregister\citealp7 
\soulregister\ref7 

\journal{International Journal of Hydrogen Energy}

\makeatletter
\def\@author#1{\g@addto@macro\elsauthors{\normalsize%
    \def\baselinestretch{1}%
    \upshape\authorsep#1\unskip\textsuperscript{%
      \ifx\@fnmark\@empty\else\unskip\sep\@fnmark\let\sep=,\fi
      \ifx\@corref\@empty\else\unskip\sep\@corref\let\sep=,\fi
      }%
    \def\authorsep{\unskip,\space}%
    \global\let\@fnmark\@empty
    \global\let\@corref\@empty  
    \global\let\sep\@empty}%
    \@eadauthor={#1}
}
\makeatother

\begin{document}

\begin{frontmatter}



\title{Strain gradient plasticity modeling of hydrogen diffusion to the crack tip}


\author{E. Mart\'{\i}nez-Pa\~neda\corref{cor1}\fnref{Uniovi}}
\ead{mail@empaneda.com}

\author{S. del Busto\fnref{Uniovi}}

\author{C. F. Niordson\fnref{DTU}}

\author{C. Beteg\'{o}n\fnref{Uniovi}}

\address[Uniovi]{Department of Construction and Manufacturing Engineering, University of Oviedo, Gij\'on 33203, Spain}

\address[DTU]{Department of Mechanical Engineering, Solid Mechanics, Technical University of Denmark, DK-2800 Kgs. Lyngby, Denmark}

\cortext[cor1]{Corresponding author. Tel: +34 985 18 19 67; fax: +34 985 18 24 33.
}

\begin{abstract}
In this work hydrogen diffusion towards the fracture process zone is examined accounting for local hardening due to geometrically necessary dislocations (GNDs) by means of strain gradient plasticity (SGP). Finite element computations are performed within the finite deformation theory to characterize the gradient-enhanced stress elevation and subsequent diffusion of hydrogen towards the crack tip. Results reveal that GNDs, absent in conventional plasticity predictions, play a fundamental role on hydrogen transport ahead of a crack. SGP estimations provide a good agreement with experimental measurements of crack tip deformation and high levels of lattice hydrogen concentration are predicted within microns to the crack tip. The important implications of the results in the understanding of hydrogen embrittlement mechanisms are thoroughly discussed.\\


\end{abstract}

\begin{keyword}

Strain gradient plasticity \sep hydrogen embrittlement \sep fracture mechanics \sep hydrogen diffusion \sep finite element analysis



\end{keyword}

\end{frontmatter}


\section{Introduction}
\label{Introduction}

When exposed to hydrogen, high-strength alloys suffer from a loss of ductility and toughness leading to premature failure \citep{S16}. The atomistic mechanism for hydrogen embrittlement remains controversial with two major candidates being inferred from experiments: hydrogen enhanced decohesion (HEDE) and hydrogen enhanced localized plasticity (HELP). Models adopting the hypothesis that interstitial hydrogen lowers the cohesive strength are able to capture the experimental trends depicted by high-strength steels in aqueous solutions and hydrogen-containing gaseous environments (see \citealp{G03}). Several attractive hydrogen-sensitive cohesive zone formulations have been proposed within this HEDE framework (e.g., \citealp{S04,S08,A13,A14}); and accurate estimations of the threshold stress intensity $K_{TH}$ and the stage II subcritical crack growth rate have been obtained by means of the \citet{G91} dislocation-based model \citep{T03,LG07,G14}. However, uncertain adjustable parameters are a shortcoming of the models and it is necessary to better define conditions within 0.1-5 $\mu$m of the crack tip, where dislocations, microstructure and chemistry dominate material behavior \citep{G05}. An accurate characterization of crack tip stresses appears fundamental as the hydrostatic stress dominates both lattice hydrogen concentration and interface decohesion.\\

The seminal paper by \citet{SM89} established the basis of hydrogen diffusion to the fracture process zone: lattice hydrogen concentration increases with distance from the crack tip, reaching its maximum at the peak site of the hydrostatic stress. The aforementioned hydrogen distribution follows the trend depicted by crack tip stresses in finite strain $J2$ plasticity and suggests that hydrogen trapped at microstructural sites plays a major role. However, classical continuum theories are unable to adequately characterize behavior at the small scales involved in crack tip deformation. Particularly, accounting for the influence of geometrically necessary dislocations (GNDs) appears imperative, as the plastic zone adjacent to the crack tip is physically small and contains strong spatial gradients of deformation.\\

As a consequence of increasing interest in microtechonology, the role of GNDs, associated with non-uniform plastic deformation, has been thoroughly investigated in a wide range of metallic materials. Thus, micro-test experiments such as bending \citep{SE98}, torsion \citep{F94} or nano-indentation \citep{NG98} have shown that metals display a strong size effect, with smaller being stronger, when non-uniform plastic deformation is confined within a small volume. In parallel, a large theoretical (see, e.g., \citealp{A84,G99,FH01,F14}) and numerical (see, e.g., \citealp{NH03,B10,K13,M15}) literature has appeared seeking to model the experimentally observed increase in yield strength and material hardening with diminishing size. In order to do so, several continuum strain gradient plasticity (SGP) theories have been developed through the years to incorporate length scale parameters in the constitutive equations. Gradient theories have been employed to provide a refined characterization of the stress distribution ahead of a crack and several authors have shown that GNDs close to the crack tip promote local strain hardening, leading to a much higher stress level as compared with classical plasticity predictions \citep{WH97,K08,N12}. Moreover, Mart\'{\i}nez-Pa\~neda and co-workers \citep{MB15,MN15} have extended these studies to the finite deformation framework, revealing a significant increase of the GND-dominated zone, as crack tip blunting is severely reduced due to the contribution of strain gradients to the work hardening of the material. Their results show that SGP predictions deviate from conventional plasticity in a physical length that spans tens of $\mu$m, highlighting the need to account for GNDs in the modelization of many damage mechanisms. As an example, traction levels estimated by SGP have been employed to justify the experimental observation of cleavage fracture in the presence of significant plastic flow \citep{E94,Q04}.\\

Although several authors (see, e.g., \citealp{T03,GS12,T15}) have noted that GNDs may be of critical relevance in hydrogen assisted cracking, its influence in hydrogen transport has not been assessed. In the present work, crack tip hydrogen diffusion is examined within a large strain framework by means of strain gradient plasticity. Several cases of particular interest are addressed with the aim of gaining insight into the role of dislocations in the continuum modeling of hydrogen diffusion. Results obtained are compared to available experimental data and physical implications are thoroughly discussed.

\section{Numerical framework}
\label{Full description}

Hydrogen diffusion to the crack tip is evaluated by means of a stress-diffusion finite element framework. A decoupled scheme is developed where a stress analysis is first conducted so as to compute the hydrostatic stress $\sigma_H$ at a certain load level. The nodal averaged value of $\sigma_H$ is then provided as initial condition in a subsequent diffusion study. Details of the finite strain SGP formulation employed in the stress computations are given in section \ref{FH01} while the diffusion analysis is described in section \ref{Diffusion analysis}.

\subsection{Strain Gradient Plasticity}
\label{FH01}

The strain gradient generalization of $J2$ flow theory proposed by \citet{FH01} is adopted to phenomenologically account for geometrically necessary dislocations in the continuum modeling. Hardening effects due to plastic strain gradients are included through the gradient of the plastic strain rate $\dot{\varepsilon}^p_{ij,k}=\left(m_{ij} \dot{\varepsilon}^p \right)_{,k}$. Where $\dot{\varepsilon}^p=\sqrt{\frac{2}{3} \dot{\varepsilon}_{ij}^p \dot{\varepsilon}_{ij}^p}$ is the increment in the conventional measure of the effective plastic strain and $m_{ij}=\frac{3}{2}s_{ij}/\sigma_e$ is the direction of the plastic strain increment, with $s_{ij}$ denoting the stress deviator, and $\sigma_e$ the von Mises effective stress. The third order plastic strain gradient tensor $\dot{\varepsilon}^p_{ij,k}$ can be decomposed, via orthogonal decomposition, into three independent tensors $\dot{\varepsilon}^{p(N)}_{ij,k}$, with $N=1,3$ \citep{SF96}. Such that a gradient-enhanced effective plastic strain rate, $\dot{E}^p$ can be defined in terms of three unique, non-negative invariants of $\dot{\varepsilon}^p_{ij,k}$, which are homogeneous of degree two:

\begin{equation}
\dot{E}_p=\sqrt{\dot{\varepsilon}^2_p + l_1^2 \dot{\varepsilon}^{p(1)}_{ij,k} \dot{\varepsilon}^{p(1)}_{ij,k} + 4 l_2^2 \dot{\varepsilon}^{p(2)}_{ij,k} \dot{\varepsilon}^{p(2)}_{ij,k}+ \frac{8}{3} l_3^2 \dot{\varepsilon}^{p(3)}_{ij,k} \dot{\varepsilon}^{p(3)}_{ij,k}}
\end{equation}

\noindent where, $l_1$, $l_2$ and $l_3$ are material length parameters. The effective plastic strain rate can be expressed explicitly in terms of $\dot{\varepsilon}^p$ and $\dot{\varepsilon}_{,i}^p$ by making use of the coefficients $A_{ij}$, $B_i$ and $C$ \citep{FH01}:

\begin{equation}
\dot{E}_p=\sqrt{\dot{\varepsilon}^2_p + A_{ij} \dot{\varepsilon}_{,i}^p \dot{\varepsilon}_{,j}^p+ B_i \dot{\varepsilon}_{,i}^p \dot{\varepsilon}^p+C \dot{\varepsilon}^{p^2}}
\end{equation}



For a body of volume $V$ and surface $S$, the principle of virtual work in the current configuration is given by

\begin{equation}\label{PVW}
\int_V \left(\sigma_{ij} \delta \dot{\varepsilon}_{ij} - \left(Q - \sigma_e \right) \delta \dot{\varepsilon}^p + \tau_i \delta \dot{\varepsilon}^p_{,i} \right) \textnormal{dV} = \int_S \left( T_i \delta \dot{u}_i + t \delta \dot{\varepsilon}^p \right)\textnormal{dS}
\end{equation}

Here, $\dot{u}_i$ is the displacement rate, $\dot{\varepsilon}_{ij}$ is the total strain rate, $\sigma_{ij}$ is the symmetric Cauchy stress tensor, $Q$ is a generalized effective stress (work conjugate to the conventional effective plastic strain) and $\tau_i$ is the higher order stress (work conjugate to the plastic strain gradient). The surface integral contains traction contributions from the conventional surface traction $T_i=\sigma_{ij} n_j$ and the higher order traction $t=\tau_i n_i$.

A finite strain version of the gradient theory by \citet{FH01} is implemented following the work of \citet{NR04}, where a thorough description can be found (see also \citealp{NT05}). An updated Lagrangian configuration is adopted and by means of Kirchhoff stress measures the incremental principle of virtual work, Eq. (\ref{PVW}), can be expressed as:

\begin{align}\label{Eq:PVW2}
& \int_V \left( \accentset{\triangledown}{\varsigma}_{ij} \delta \dot{\varepsilon}_{ij} - \sigma_{ij} \left(2 \dot{\varepsilon}_{ik} \delta \dot{\varepsilon}_{kj} - \dot{e}_{kj} \delta \dot{e}_{ki} \right)+ \left(\dot{q} - \dot{\sigma}_e^{\varsigma} \right)\delta \dot{\varepsilon}^p + \accentset{\vee}{\rho}_{i} \delta \dot{\varepsilon}_{,i}^p \right) \textnormal{dV} \nonumber \\
&= \int_S \left( \dot{T}_{0i} \delta \dot{u}_i + \dot{t}_0 \delta \dot{\varepsilon}^p \right)\textnormal{dS}
\end{align}

Here, $\accentset{\triangledown}{\varsigma}_{ij}$ is the Jaumann rate of the conventional Kirchhoff stress, $\dot{q}$ is the rate of the Kirchhoff variant of the effective stress, $\accentset{\vee}{\rho}_{i}$ is the convected derivative of the higher order Kirchhoff stress and the velocity gradient is denoted by $\dot{e}_{ij}$. The Kirchhoff quantities are related to their Cauchy counterparts in Eq. (\ref{PVW}) by the determinant, $J$, of the deformation gradient: $\varsigma_{ij}=J \sigma_{ij}$, $\rho_i=J \tau_i$, $q=JQ$ and $\sigma_e^{\varsigma}=J \sigma_e$. The incremental constitutive equations for the stress measures are given, in terms of the hardening modulus $h \left[ E_p \right]$, by:

\begin{equation}
\accentset{\triangledown}{\varsigma}_{ij} = \mathscr{D}_{ijkl} \left(\dot{\varepsilon}_{kl} - \dot{\varepsilon}^P m_{kl} \right)=\dot{\varsigma}_{ij} - \dot{\omega}_{ik} \sigma_{kj} - \sigma_{ik} \dot{\omega}_{jk}
\end{equation}

\begin{equation}
\dot{q}-\dot{\sigma}^{\varsigma}_{(e)}=h \left( \dot{\varepsilon}^P + \frac{1}{2} B_i \dot{\varepsilon}_{,i}^P +C \dot{\varepsilon}^P \right) - m_{ij} \accentset{\triangledown}{\varsigma}_{ij}
\end{equation}

\begin{equation}
\accentset{\vee}{\rho}_{i}=h \left( A_{ij} \dot{\varepsilon}_{,j}^P + \frac{1}{2} B_i \dot{\varepsilon}^P \right)= \dot{\rho}_i - \dot{e}_{ik} \rho_k
\end{equation}

\noindent where, for a given Young's modulus $E$ and Poisson ratio $\nu$, the elastic stiffness tensor equals

\begin{equation}
\mathscr{D}_{ijkl}=\frac{E}{1+\nu} \left( \frac{1}{2} \left( \delta_{ik} \delta_{jl} + \delta_{il} \delta_{jk} \right) + \frac{\nu}{1-2\nu} \delta_{ij} \delta_{kl} \right)
\end{equation}

\noindent and $\dot{\omega}_{ij}=\frac{1}{2} \left(\dot{e}_{ij}- \dot{e}_{ji} \right)$ is the anti-symmetric part of the velocity gradient.\\

A special kind of finite element (FE) method is used where, in addition to the nodal displacement increments, $\dot{D}^n$, the nodal effective plastic strain increments, $\dot{\varepsilon}^p_n$, appear directly as unknowns. The displacement increments, $\dot{u}_i$, and the effective plastic strain increments, $\dot{\varepsilon}^p$, are interpolated within each element by means of the shape functions $N_i^n$ and $M^n$:

\begin{equation}\label{Eq:FEinter}
\dot{u}_i= \sum_{n=1}^{k} N_i^n \dot{D}^n \:, \; \; \; \; \; \; \; \dot{\varepsilon}^p=\sum_{n=1}^{l} M^n \dot{\varepsilon}^p_n
\end{equation}

\noindent where $k$ and $l$ are the number of nodes used for the displacement and the effective plastic strain interpolations, respectively. Quadratic shape functions are used for the displacement field ($k=8$) while linear shape functions are employed for the effective plastic strain field ($l=4$). By introducing the FE interpolation of the displacement field and the effective plastic strain field (\ref{Eq:FEinter}), and their appropriate derivatives, in the principle of virtual work (\ref{Eq:PVW2}), the following discretized system of equations is obtained:

\begin{equation}\label{Eq:Dis}
\begin{bmatrix}
  \boldsymbol{K_e} & \boldsymbol{K_{ep}} \\
  \boldsymbol{K_{ep}^T} & \boldsymbol{K_p}
 \end{bmatrix}
\begin{bmatrix}
\dot{\boldsymbol{D}} \\
\dot{\boldsymbol{\varepsilon}}^{\boldsymbol{p}} 
\end{bmatrix}=
\begin{bmatrix}
\dot{\boldsymbol{F}}_{\boldsymbol{1}}\\
\dot{\boldsymbol{F}}_{\boldsymbol{2}}
\end{bmatrix}
\end{equation}

Here, $\boldsymbol{K_e}$ is the elastic stiffness matrix, $\boldsymbol{K_{ep}}$ is a coupling matrix of dimension force and $\boldsymbol{K_p}$ is the plastic resistance, a matrix of dimension energy. The first part of the right-hand side of Eq. (\ref{Eq:Dis}) is composed of the conventional external incremental force vector $\dot{\boldsymbol{F}}_{\boldsymbol{1}}$ and the incremental higher order force vector $\dot{\boldsymbol{F}}_{\boldsymbol{2}}$. The contribution of $\boldsymbol{K_p}$ is minimized in the elastic regime, non-restricting the plastic strain at the evolving elastic-plastic boundary as within conventional plasticity (see \citealp{NH03}).\\

Based on a forward Euler scheme, when nodal displacement and effective plastic strain increments have been determined, the updated strains, $\varepsilon_{ij}$, stresses, $\sigma_{ij}$, higher order stresses, $\tau_i$, and $Q$ are computed at each integration point.\\

\subsection{Hydrogen diffusion}
\label{Diffusion analysis}

The diffusion problem, in a volume $V$ of surface $S$ and outward normal $n_i$, is derived from the requirement of mass conservation for the diffusing phase: (see, e.g., \citealp{SM89})

\begin{equation}
\frac{\textnormal{d}}{\textnormal{d}t} \int_V c \,\, \textnormal{d}V + \int_S n_i J_i \,\, \textnormal{d} S=0
\end{equation}

Where $d/dt$ is the time derivative, $c$ is the mass concentration of the diffusing material and $J_i$ is the flux of concentration of the diffusing phase. A normalized concentration is defined $\phi=c/s$ denoting the relation between the mass concentration of the diffusing material $c$ and its solubility in the base material $s$. Within this framework, stress-driven hydrogen diffusion to the crack tip is modeled by an extended form of Fick's law: 

\begin{equation}\label{Eq:flux}
J_i=-s D \nabla \left( \phi - \kappa_p \sigma_H \right)
\end{equation}

With $D$ being the hydrogen diffusion coefficient and $\kappa_p$ the pressure stress factor, which is defined by

\begin{equation}
\kappa_p=\frac{\bar{V}_H \phi}{R \left(T - T^z \right)}
\end{equation}

Here, $\bar{V}_H$ is the partial molar volume of hydrogen, $T$ is the temperature (with $T^z=0 \, K$ being its absolute zero value) and $R$ is the universal gas constant. Time integration in the transient diffusion computations conducted is performed by means of the backward Euler method. Under steady-state conditions, the normalized concentration $\phi$ is related to the hydrostatic stress $\sigma_H$ by:

\begin{equation}\label{Eq:BC}
\phi=\phi_0 \textnormal{exp} \left( \frac{\bar{V}_H \sigma_H}{R (T-T^z)} \right)
\end{equation}

With $\phi_0$ being the normalized hydrogen concentration in the unstressed state.

\section{Finite element results}
\label{FE Results}

The role of strain gradients in hydrogen diffusion is assessed by addressing several cases of particular interest. Thus, section \ref{SofronisMcMeeking} aims to model hydrogen transport towards a blunted crack tip in iron, following the pioneering work by \citet{SM89}. While, inspired by the work by \citet{O09}, section \ref{Olden} is devoted to the modelization of crack tip blunting and transient hydrogen diffusion on duplex stainless steel. And in section \ref{X80} the distribution of hydrogen ahead of a crack in X80 pipeline steel is examined and compared to the experimental results of \citet{ML98}.

\subsection{Hydrogen transport in impure iron}
\label{SofronisMcMeeking}

\begin{figure}[H]
\centering
\includegraphics[scale=0.6]{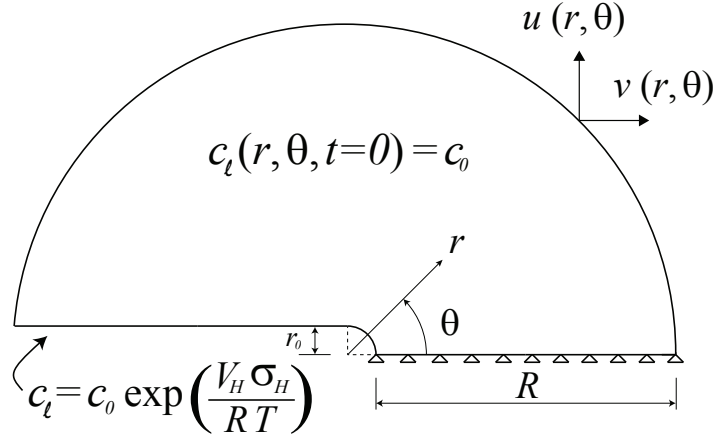}
\caption{Description of the boundary and initial conditions for the stress and diffusion model employed for the impure iron case.}
\label{fig:BCs}
\end{figure}

In their pioneering work, \citet{SM89} established the basis for hydrogen transport ahead of the crack under large strains. The influence of GNDs is first examined by mimicking their conventional plasticity calculations. Crack tip fields are evaluated in the stress analysis by means of a boundary layer formulation. Hence, as described in Fig. \ref{fig:BCs}, the crack region is contained by a circular zone and a remote Mode I load is applied by prescribing the displacements of the nodes at the remote circular boundary:

\begin{equation}
u(r,\theta)=K_I \frac{1+\nu}{E} \sqrt{\frac{r}{2\pi}}cos\left(\frac{\theta}{2}\right)(3-4\nu-cos\theta)
\end{equation}

\begin{equation}
v(r,\theta)=K_I \frac{1+\nu}{E} \sqrt{\frac{r}{2\pi}}sin\left(\frac{\theta}{2}\right)(3-4\nu-cos\theta)
\end{equation}

\begin{figure}[H]
        \centering
        \begin{subfigure}[h]{0.45\textwidth}
                \centering
                \includegraphics{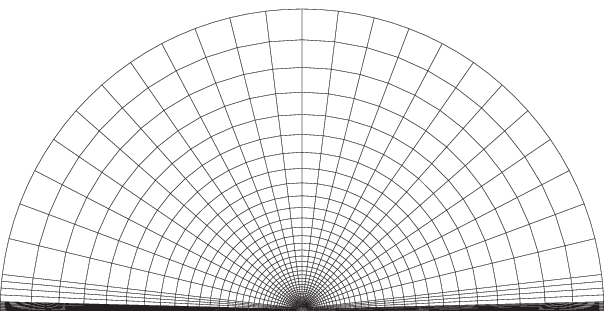}
                \caption{}
                \label{fig:FigMBLa}
        \end{subfigure}
        \begin{subfigure}[h]{0.45\textwidth}
                \centering
                \includegraphics{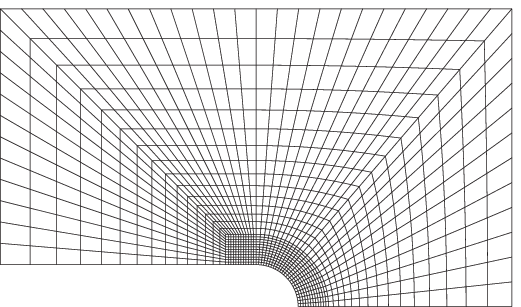}
                \caption{}
                \label{fig:FigMBLb}
        \end{subfigure}
       
        \caption{Finite element mesh: (a) complete model and (b) vicinity of the crack}\label{fig:Mesh}
\end{figure}

Here, $u$ and $v$ are the horizontal and vertical components of the displacement boundary condition, $r$ and $\theta$ the radial and angular coordinates of each boundary node in a polar coordinate system centered at the crack tip, and $K_I$ is the applied stress intensity factor, which quantifies the remote load. Following \citep{M77,SM89}, a ratio between the radii of the outer boundary and the blunted crack tip of $R/r_0=10^5$ is adopted (with $r_0=0.0005$ mm). Plane strain and small scale yielding conditions are assumed and only the upper half of the circular domain is modeled due to symmetry. After a thorough sensitivity study, a mesh of 6200 eight-noded quadrilateral elements with reduced integration is employed in both the diffusion and the stress analyses (see Fig. \ref{fig:Mesh}).\\

Regarding the diffusion model, an initial bulk concentration $c_0$ may be defined to avoid numerical oscillations (see \citealp{SM89}). The boundary concentration is prescribed in the crack flank as a function of the initial concentration and the hydrostatic stress (see Eq. (\ref{Eq:BC}) and Fig. \ref{fig:BCs}). The boundary conditions adopted accurately capture the diffusion of hydrogen to the fracture process zone under both internal and environmental assisted hydrogen cracking. Other combinations of hydrogen flux boundary conditions have been considered but, as already noted by \citet{SM89}, the sensitivity of the hydrogen distribution ahead of the crack tip is negligible. The boundary conditions employed significantly alleviate convergence problems derived from the existing steep concentration gradients and follow the concept of prescribing a constant lattice chemical potential rather than a constant lattice hydrogen concentration, as introduced by \citet{DA13}. Unlike the gradient-enhanced stress computations, the diffusion study can be easily performed in commercial FE packages as it does not require a special FE formulation. In the case of the well-known FE code ABAQUS, the outcome of the stress analysis (averaged nodal values of $\sigma_H$) can be read from a file and subsequently introduced as input in the diffusion study by means of a UPRESS subroutine. A constant lattice chemical potential can be prescribed by reading the same file within a DISP subroutine.

Results are obtained for impure iron, with its uniaxial stress-strain law being characterized by (see \citealp{SM89})

\begin{equation}
\left( \frac{\sigma}{\sigma_Y} \right)^{1/N} = \frac{\sigma}{\sigma_Y}+\frac{3 \mu}{\sigma_Y} \varepsilon^p
\end{equation}

\begin{figure}[H]
\centering
\includegraphics[scale=1.15]{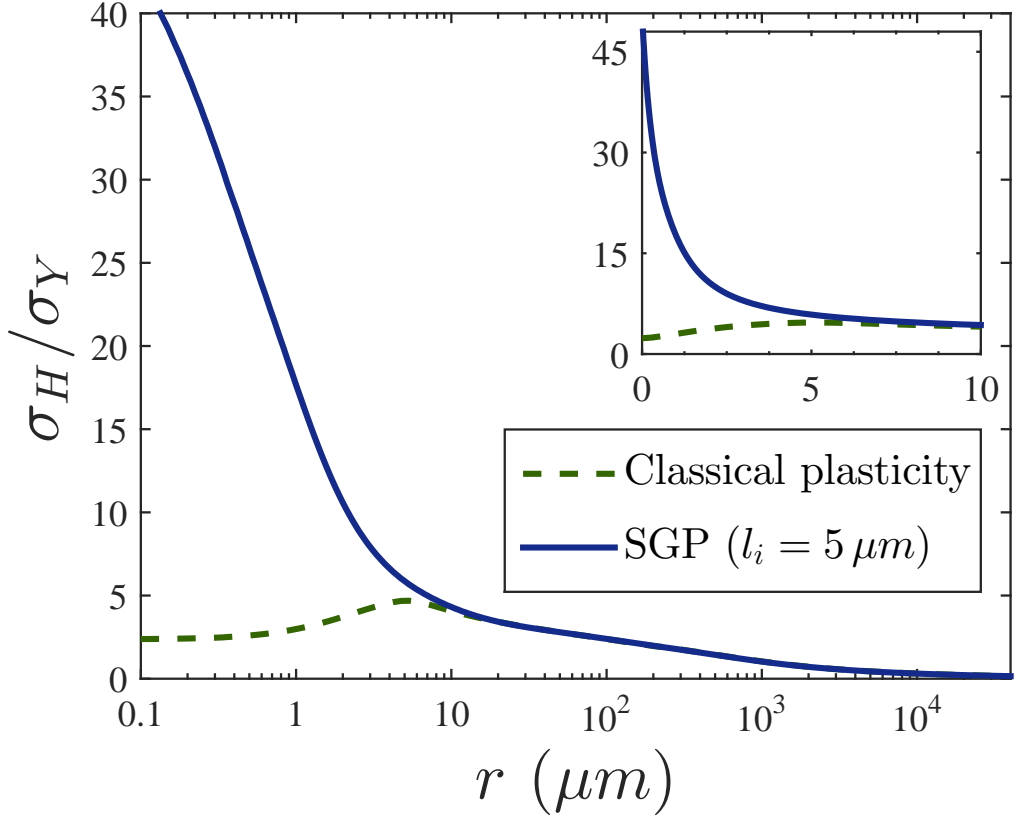}
\caption{Normalized hydrostatic stress distribution ahead of the crack tip in impure iron for an external load of $K_I= 89.7$ MPa$\sqrt{m}$ from SGP (with $l_i=5 \, \mu m$) and classical plasticity. The figure shows results along the extended crack plane with the distance to the crack tip in linear (inset) and logarithmic (main figure) scales.}
\label{fig:SHiron}
\end{figure}

Where the strain hardening exponent, the yield stress and the shear modulus are given by $N=0.2$, $\sigma_Y=250$ MPa and $\mu=79.6$ GPa ($E=207$ GPa and $\nu=0.3$), respectively. A material length scale of $l_1=l_2=l_3=5 \, \mu m$ is adopted in the gradient-enhanced computations. This would be a typical estimate for nickel (see \citealp{SE98}) and corresponds to an intermediate value within the range of experimentally fitted length scales reported in the literature. The hydrostatic stress distribution obtained for an external load of $K_I= 89.7$ MPa$\sqrt{m}$ is shown in Fig. \ref{fig:SHiron}. The stress values are normalized by $\sigma_Y$ while the distance to the crack tip $r$ is left unchanged, with the aim of assessing the physical length where strain gradients are particularly relevant. As SGP theories describe the collective behavior of a significant number of dislocations, they are only applicable at a scale much larger than the average dislocation spacing. For common values of dislocation density in metals, a lower limit of physical validity could be established around 100 nm and consequently results are generally shown beyond the aforementioned distance to the crack tip. The abscissa axis is plotted in logarithmic scale for the sake of clarity, but results in a regular linear scale are also shown in the inset of the figure.

\begin{figure}[H]
\centering
\includegraphics[scale=1.1]{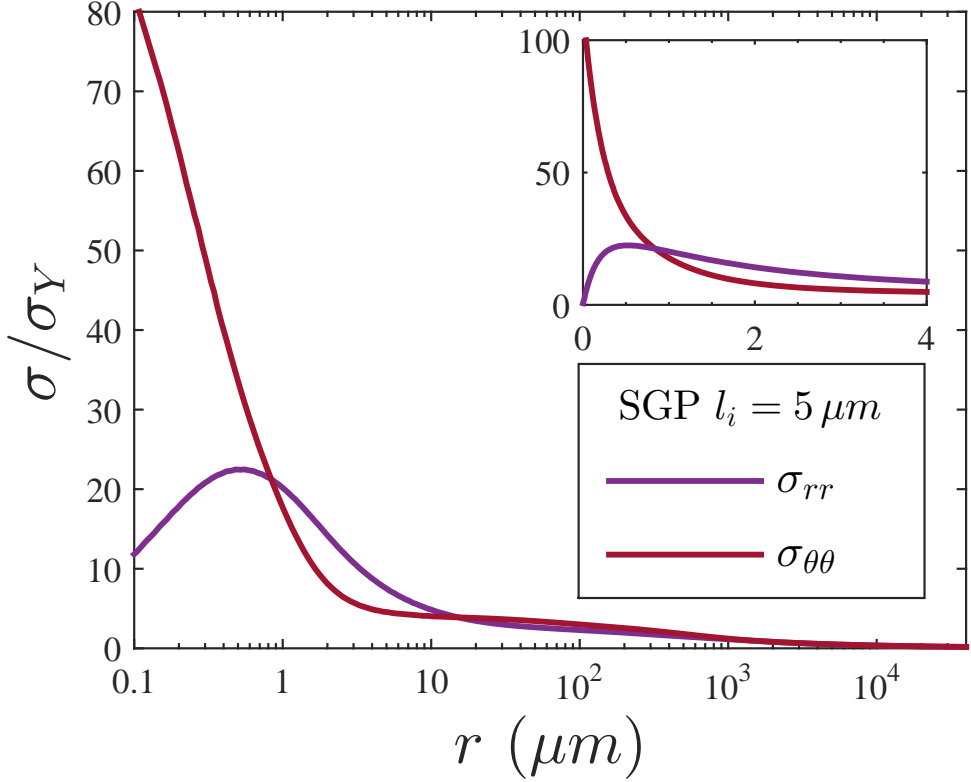}
\caption{Normalized opening ($\sigma_{\theta \theta}$) and axial ($\sigma_{r r}$) stress distributions ahead of the crack tip in impure iron for an external load of $K_I= 89.7$ MPa$\sqrt{m}$ from SGP (with $l_i=5 \, \mu m$). The figure shows results along the extended crack plane with the distance to the crack tip in linear (inset) and logarithmic (main figure) scales.}
\label{fig:SHiron2}
\end{figure}

Classical plasticity predictions reproduce the well known behavior revealed by \citet{M77}: $\sigma_H$ increases for decreasing values of $r$ until the stresses become influenced by crack blunting. As large strains cause the crack to blunt, the stress triaxiality is reduced locally, with $\sigma_H$ reaching a peak at - in the present case study - $r \approx 6 \, \mu m$. However, a monotonic increase of the stress level is observed when the influence of strain gradients is accounted for. SGP predictions agree with $J2$ plasticity far from the crack tip but significant differences arise in the vicinity of the crack, as the density of GNDs increases. Gradient plasticity estimations of $\sigma_H$ are based on the trend depicted by the opening ($\sigma_{\theta \theta}$) and axial stresses ($\sigma_{r r}$), as shown in Fig. \ref{fig:SHiron2}. $\sigma_{\theta \theta}$ increases monotonically as the distance to the crack tip decreases while $\sigma_{r r}$ vanishes at the free surface.
 
Local stress reduction does not take place in SGP due to the contribution of strain gradients to the work hardening of the material \citep{MB15,MN15}. The influence on hydrogen diffusion of the macroscopic stress elevation attained due to gradient-enhanced hardening is subsequently examined.\\

\begin{figure}[H]
\centering
\includegraphics[scale=1.1]{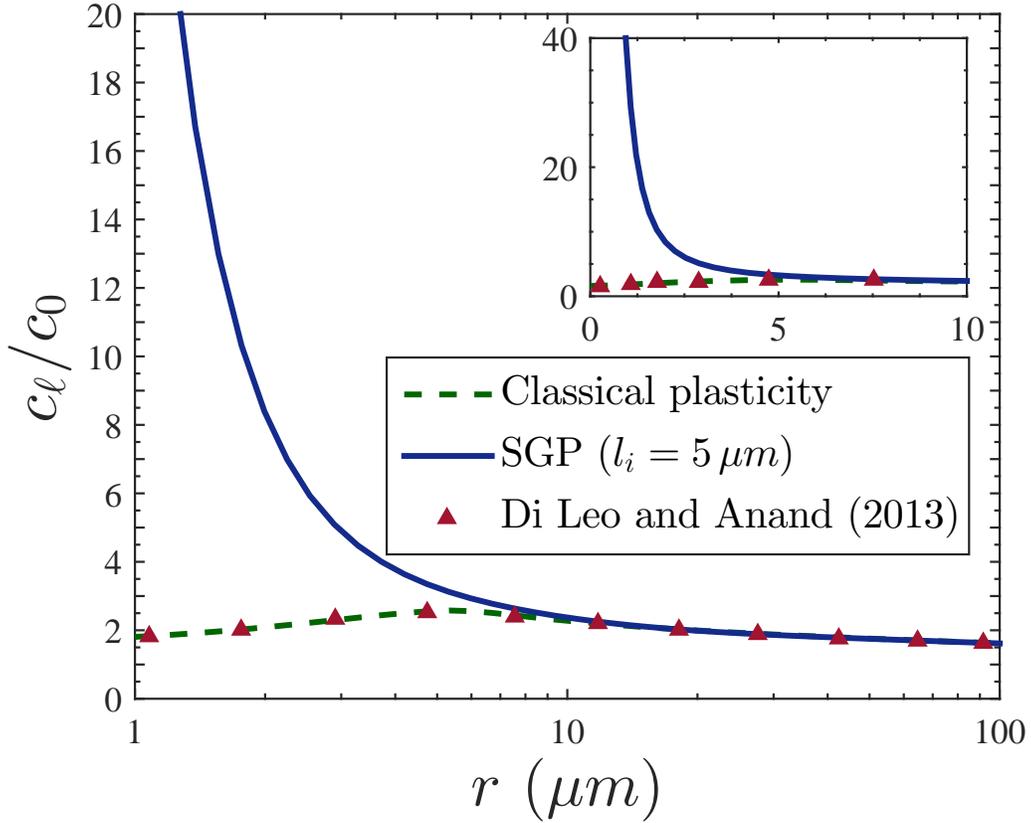}
\caption{Normalized concentration of lattice hydrogen ahead of the crack tip in impure iron for SGP (with $l_i=5 \, \mu m$) and classical plasticity. The figure shows results along the extended crack plane after 1419 hours, with the distance to the crack tip in linear (inset) and logarithmic (main figure) scales.}
\label{fig:Ciron}
\end{figure}

Fig. \ref{fig:Ciron} shows the results obtained in the subsequent diffusion study of hydrogen transport in impure iron. Following \citep{SM89}, the lattice diffusion constant is given by $D=1.27 \cdot 10^{-8} \, m^2 s^{-1}$ and the initial concentration of hydrogen in the bulk is $c_0=2.084 \cdot 10^{21}$ atoms per $m^3$. As in \citep{SM89,DA13}, the distribution of lattice hydrogen concentration $c_{\ell}$ ahead of the crack tip is computed after 1419 hours.

As discussed in \citep{DA13}, after 500 hours the change in concentration rate is negligible, such that $t=1419$ h is well beyond the time at which steady-state conditions are first reached. By prescribing a constant lattice chemical potential - as opposed to a constant lattice hydrogen concentration - numerical predictions are able to match the steady-state profile predicted by  Eq. (\ref{Eq:BC}).\\

As in \citep{DA13}, the distribution of lattice hydrogen estimated by means of conventional plasticity reaches a peak at $c_{\ell}/c_0 \approx 2.7$ and then decreases as the crack tip is approached. On the contrary, in accordance with the trend depicted by $\sigma_H$, when strain gradients are accounted for the hydrogen concentration increases monotonically as $r$ decreases. Consequently, significant differences arise between the predictions of conventional and gradient-enhanced plasticity formulations, with the latter estimating high levels of lattice hydrogen close to the crack surface. Results reveal that GNDs, absent in conventional plasticity predictions, play a very relevant role in hydrogen diffusion ahead of a crack tip.

\subsection{Crack tip blunting and hydrogen distribution in duplex stainless steel}
\label{Olden}

\begin{figure}[H]
\makebox[\linewidth][c]{%
        \begin{subfigure}[b]{0.8\textwidth}
                \centering
                \includegraphics[scale=0.5]{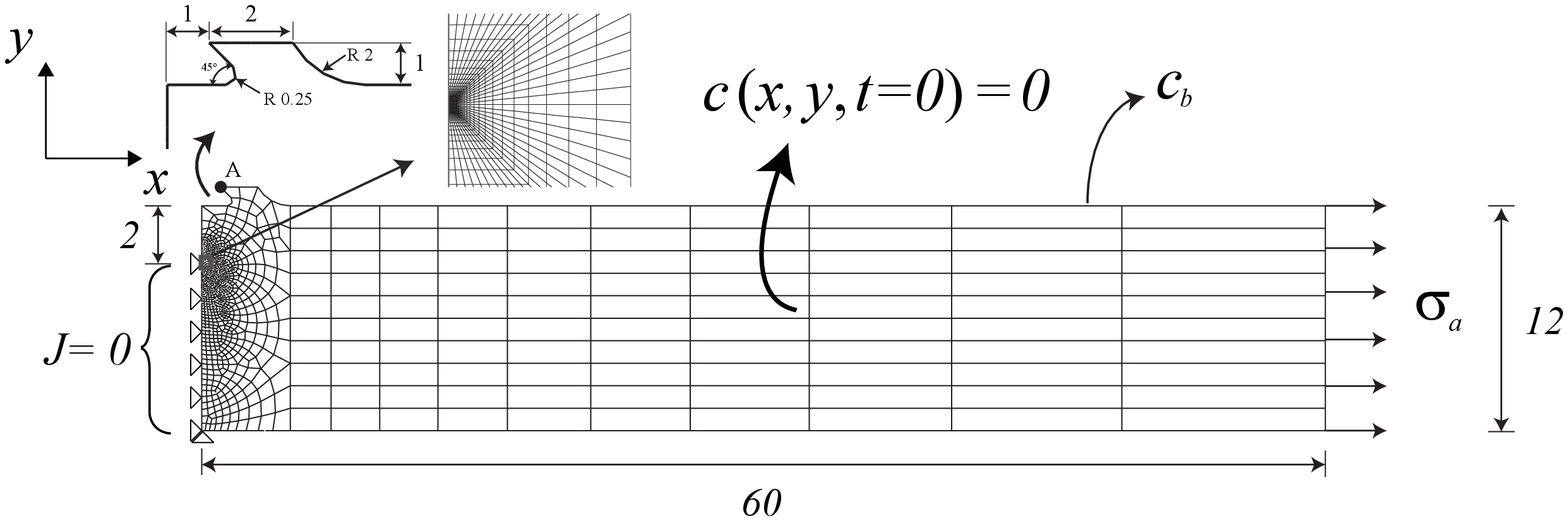}
                \caption{}
                \label{fig:Config}
        \end{subfigure}
        \begin{subfigure}[b]{0.55\textwidth}
                \raggedleft
                \includegraphics[scale=0.37]{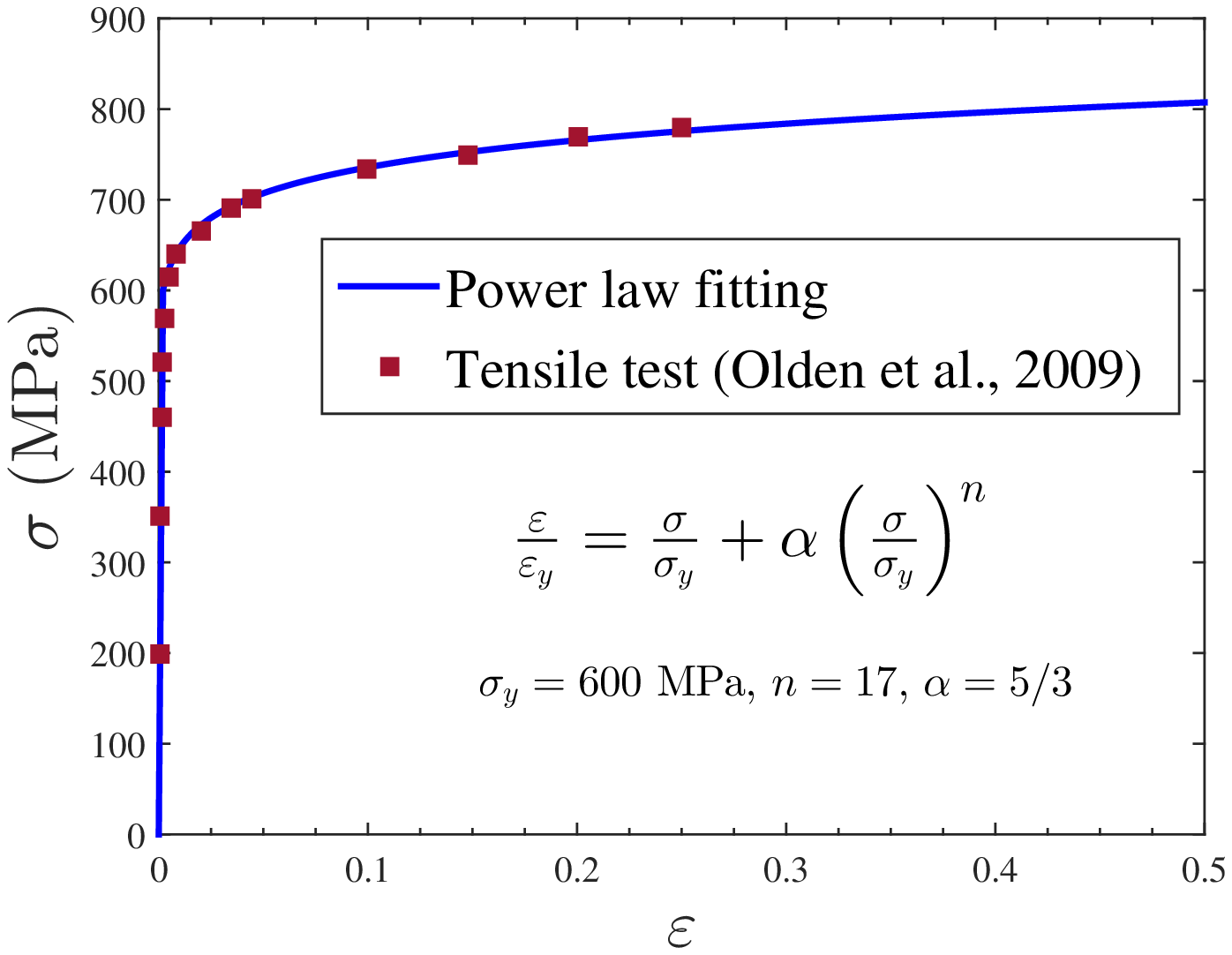}
                \caption{}
                \label{fig:SScurve}
        \end{subfigure}
        }
       
        \caption{Finite element model for the duplex stainless steel study: (a) Mesh, geometry and boundary conditions, with all dimensions in mm and (b) Stress-strain curve}\label{fig:DuplexSteel}
\end{figure}

Despite its wide use in sub-sea applications, duplex stainless steels are sensitive to environmentally assisted hydrogen cracking at low corrosion protection potentials \citep{O08}. The role of plastic strain gradients on the onset of damage in 25\%Cr duplex stainless steel is assessed by estimating the distribution of lattice hydrogen in the experiments carried out by \citet{O09}. Hence, single edge notched tensile (SENT) specimens under constant load and cathodic protection are examined. Due to symmetry, only half of the SENT specimen is modeled, as depicted in Fig. \ref{fig:Config}. A mesh of approximately 4000 quadratic quadrilateral plane strain elements is employed, with an element size of a few nanometers in the vicinity of the 2 mm fatigued pre-crack. The material behavior is characterized by fitting the stress-strain tensile test data shown in Fig. \ref{fig:SScurve}, and the load is applied by imposing an applied stress $\sigma_a$ in the right edge of the specimen.

\begin{figure}[H]
\centering
\includegraphics[scale=0.95]{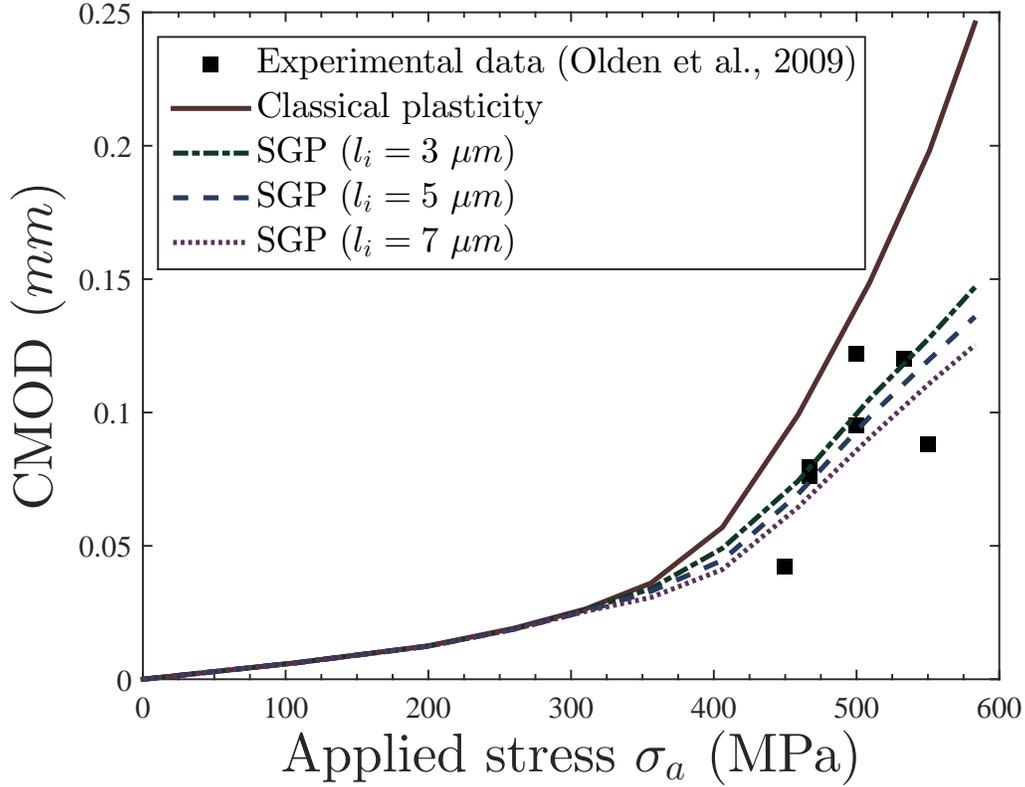}
\caption{Experimental and numerical predictions for the crack mouth opening displacement in duplex stainless steel.}
\label{fig:CMOD}
\end{figure}

The stress analysis leads to the qualitative output depicted in Section \ref{SofronisMcMeeking}: a monotonic increase of the hydrostatic stress is observed when strain gradients are accounted for. As outlined before, the stress triaxiality reduction near the crack tip intrinsic to classical plasticity is not observed in SGP theories due to the contribution of the strain gradients to the work hardening of the material. Namely, enhanced dislocation hardening significantly lowers crack tip blunting with respect to conventional plasticity predictions (see \citealp{MN15}). Fig. \ref{fig:CMOD} shows the crack mouth opening displacement (CMOD, measured at point A in Fig. \ref{fig:Config}) computed for several load levels from both classical plasticity and SGP. Computations are performed for three values of the intrinsic material length $l_i$ with the aim of assessing the role of the parameter(s) governing the influence of the GNDs density. The experimentally measured data of \citet{O09} is also included. 

As shown in Fig. \ref{fig:CMOD}, crack blunting is significantly reduced when GNDs are accounted for, with the differences with classical plasticity increasing with the load. Results also show little influence of the material length scale $l_i$, despite varying its value over the range of experimentally reported values. Moreover, SGP predictions seem to provide a better fit with the experiments of \citet{O09}.\\

\begin{figure}[H]
\centering
\includegraphics[scale=0.9]{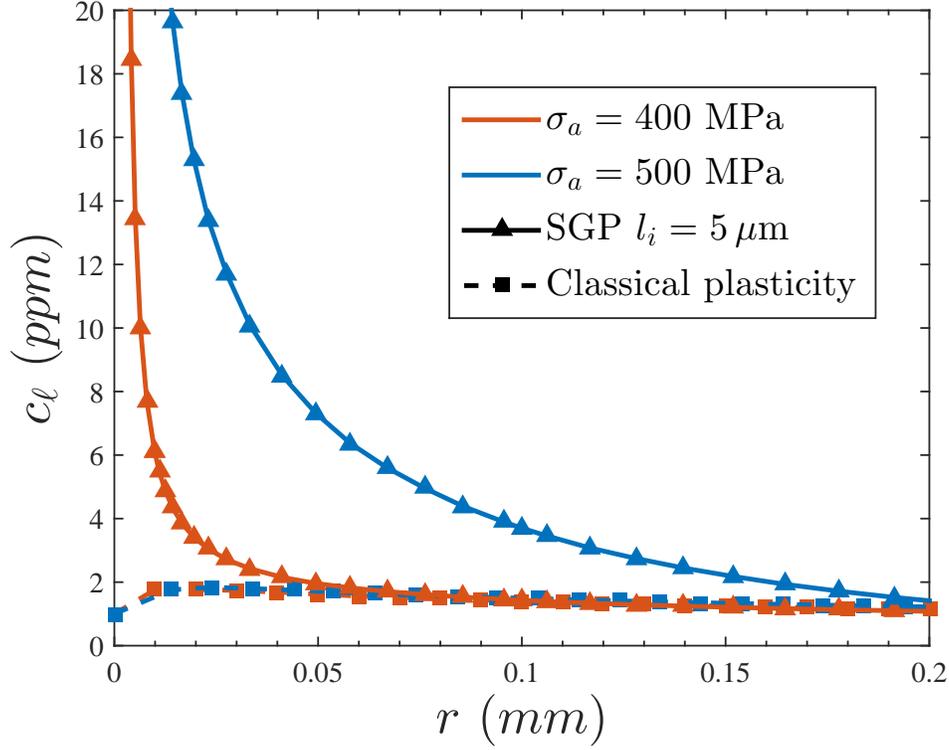}
\caption{Concentration of lattice hydrogen ahead of the crack tip in duplex stainless steel for SGP (with $l_i=5 \, \mu m$) and classical plasticity. The figure shows results along the extended crack plane for different applied stresses $\sigma_a$ after 200 h.}
\label{fig:Cduplex}
\end{figure}

For the subsequent diffusion study two different load levels have been considered, which correspond to net section stresses of 480 and 600 MPa or, equivalently, 80-100 \% of the material yield strength (typical service stress levels for sub-sea pipelines are in the range of 60-80\% of the yield stress). Following \citet{O09}, a surface hydrogen concentration of 1 ppm is assumed, which corresponds to the conditions of the experimental setup (3\% NaCl solution, artificial sea water at 4$^{\circ}$C and an applied cathodic potential of -1050 $mV_{SCE}$). A transient study is conducted with the aim of assessing crack tip hydrogen concentration after 200 h. of exposure. The diffusion coefficient is estimated to be $3.7 \cdot 10^{-12}$ $m^2/s$ (see \citealp{O09}). Boundary conditions are depicted in Fig. \ref{fig:Config} with a constant lattice hydrogen concentration being prescribed, unlike the previous case study. Since GNDs lead to steep concentration gradients and a surface hydrogen concentration is imposed at the crack flanks, numerical convergence (with a negligible effect in the results for $r>0.1 \, \mu$m) can be significantly improved by isolating (i.e., $J=0$) the node at the crack tip. Results are shown in Fig. \ref{fig:Cduplex}, where the SGP estimations have been computed for the intermediate value of the material length parameter ($l_i=5 \, \mu$m).\\

Results reveal a major influence of GNDs over physically meaningful distances, with the lattice hydrogen concentration predicted by means of SGP significantly increasing within 0.05-0.1 mm to the crack tip. Classical plasticity predictions, in agreement with the computations of \citet{O09}, show little sensitivity to the external load. This is not the case if strain gradients are accounted for, as the lattice hydrogen level increases with the applied stress. 

\subsection{Crack tip hydrogen concentration in X80 pipeline steel}
\label{X80}

\begin{figure}[H]
\centering
\includegraphics[scale=0.9]{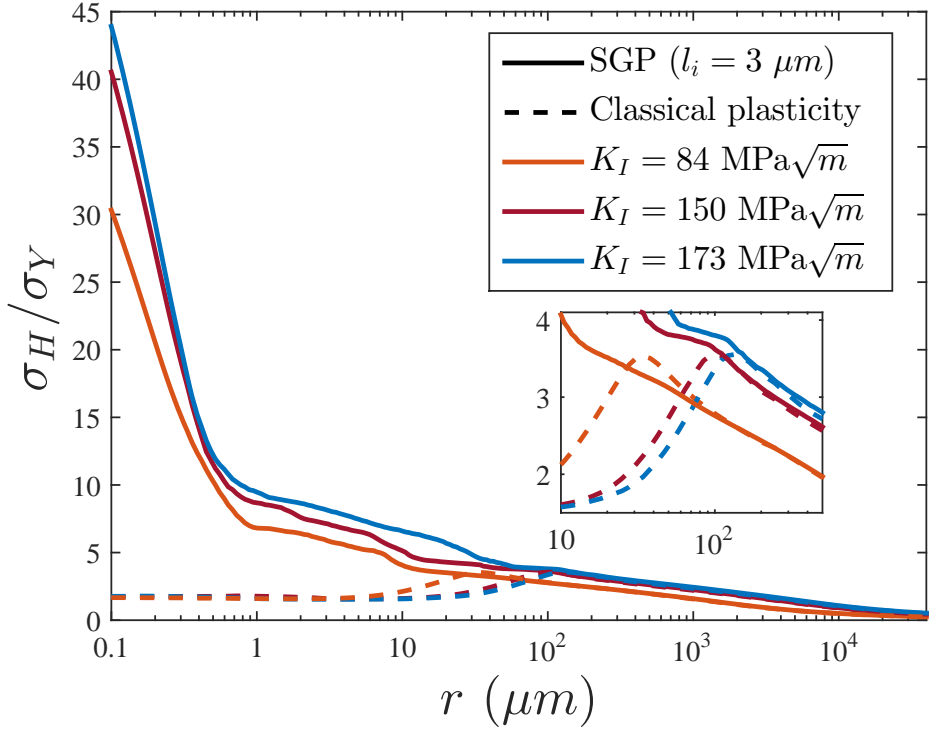}
\caption{Normalized hydrostatic stress distribution ahead of the crack tip in X80 pipeline steel for SGP (with $l_i=5 \, \mu m$) and classical plasticity. The figure shows results ahead of the crack tip for different load levels with the distance to the crack tip in log scale.}
\label{fig:SHx80}
\end{figure}

There is a strong consensus that large gradients of plastic strain close to the crack tip promote additional hardening and very high crack tip stresses that classical plasticity is unable to capture. This must undoubtedly lead to high concentration of lattice hydrogen close to the crack surface. However, an experimental quantitative assessment is complicated as differences are located within a physical length on the order of micrometers. Secondary ion mass spectrometry (SIMS) seems to be one of the few techniques able to accurately measure hydrogen concentration profiles at such scales. By means of SIMS, \citet{ML98} were able to measure the hydrogen distribution around a crack tip in X80 pipeline steel. In their experimental work, compact tension specimens were first loaded in the absence of hydrogen and then immersed in NS-4 solution at free potential for 72 h (typical test solution for coating disbondment in Canadian pipelines, more details can be found in \citealp{ML98}). Their experimental setup is modeled with the aim of gaining quantitative insight into the role of GNDs in crack tip hydrogen diffusion. As in section \ref{SofronisMcMeeking} the remote mode I load is imposed by means of a boundary layer formulation with three load levels being considered (see \citealp{ML98}): $K_I= 84$ MPa$\sqrt{m}$ ($\textnormal{J}=32 \cdot 10^3 \, J/m^2$), $K_I= 150$ MPa$\sqrt{m}$ ($\textnormal{J}=102 \cdot 10^3 \, J/m^2$) and $K_I= 173$ MPa$\sqrt{m}$ ($\textnormal{J}=136 \cdot 10^3 \, J/m^2$). The elastic parameters of X80 steel are $E=200$ GPa and $\nu=0.3$. A yield stress of $\sigma_Y=600$ MPa is adopted and following \citet{ML98} a hardening law of the type

\begin{equation}
\frac{\varepsilon}{\varepsilon_y}=\frac{\sigma}{\sigma_Y}+\alpha \left(\frac{\sigma}{\sigma_Y} \right)^n
\end{equation}

\noindent is assumed, with $\varepsilon_y$ being the yield strain ($\sigma_Y/E$). The dimensionless constant $\alpha$ and the strain hardening exponent $n$ respectively adopt the values of 0.01 and 6.6, respectively. A value of $l_i=3 \, \mu m$ is adopted within the SGP formulation. The choice is based on the good agreement observed with the CMOD measurements in 25\%Cr duplex stainless steel for $l_i=5 \, \mu m$ and the fact that a higher degree of work hardening may be associated with a lower value of $l_i$ (see, e.g., the expression for $l$ provided by MSG plasticity, \citealp{Q04,MB15}). Fig. \ref{fig:SHx80} shows the hydrostatic stress distribution computed ahead of the crack tip for both classical and strain gradient plasticity formulations. 

Results show that significantly higher stress levels are attained with the SGP formulation. The differences with respect to classical plasticity predictions are relevant in a domain that spans several tens of $\mu$m, embracing the critical distance of many damage mechanisms. Fig. \ref{fig:SHx80} also shows a distinct feature of conventional plasticity: the value of the peak stress remains constant as the applied load increases, while its location moves away from the crack tip. This peculiarity of large strain $J2$ plasticity - on which many damage models are based - is not observed when GNDs are constitutively involved. On the contrary, the degree of stress elevation attained by means of SGP increases with the external load. Thereby, results reveal great differences between gradient and classical plasticity as the load increases, with $\sigma_H$ in the former being more than 20 times the conventional prediction.\\

\begin{figure}[H]
\centering
\includegraphics[width=1.2\textwidth,left]{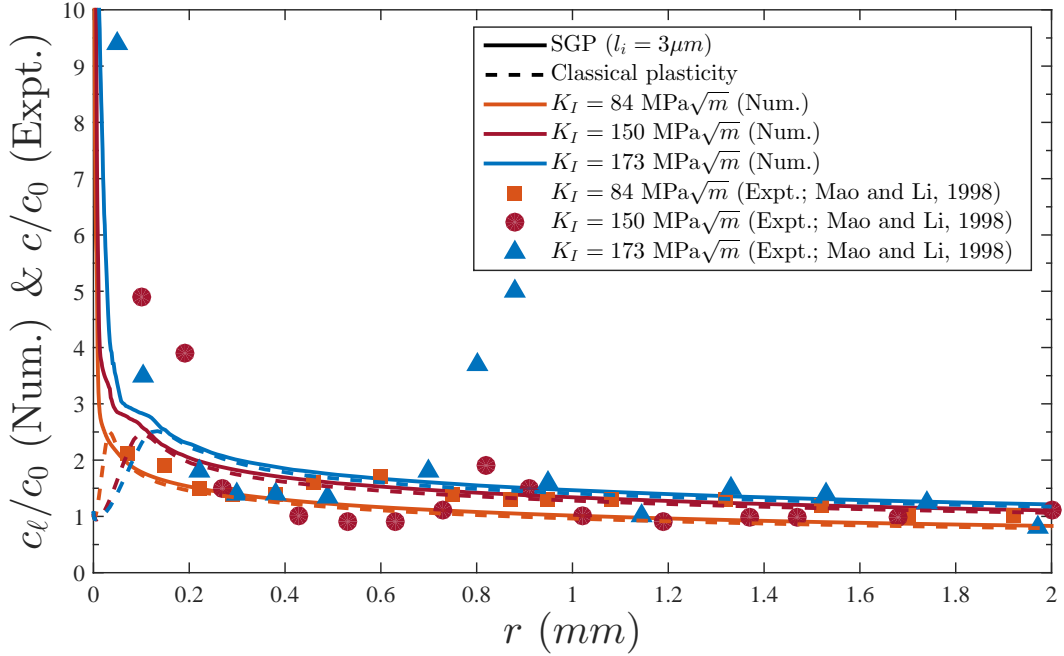}
\caption{Experimental measurements and numerical predictions of, respectively, the diffusible and lattice hydrogen concentrations ahead of the crack tip in X80 pipeline steel after 72 h. The figure shows results along the extended crack plane for different load levels.}
\label{fig:CX80}
\end{figure}

A subsequent diffusion study is conducted where, mimicking the experimental setup, a bulk initial concentration of $c_{\ell}(t=0)=0$ is defined and a boundary concentration of $c_{\ell}=c_0$ is imposed on crack flanks and outer radius. Convergence issues due to steep gradients can be alleviated by isolating a few nodes close to the crack tip, as in the previous case study. A lattice diffusion constant of $D=6.699 \cdot 10^{-11} \, m^2 s^{-1}$ is adopted, following the experimental measurements by \citet{H14}. The numerical results obtained after 72 h for both classical and gradient-enhanced formulations are shown in Fig. \ref{fig:CX80}. The experimental SIMS measurements performed by \citet{ML98} are also included.

Fig. \ref{fig:CX80} reveals important differences between conventional and gradient-enhanced predictions, with very high values of lattice hydrogen being predicted in the vicinity of the crack if certain microstructural features (GNDs) are incorporated into the modeling. The same trends are observed for the experimental measurements of diffusible hydrogen $c$ and SGP-based predictions of $c_{\ell}$; namely, hydrogen concentration (i) increases with the external load and (ii) raises sharply within microns to the crack tip as $r \to 0$.

\subsection{The role of hydrogen trapping}
\label{Trapping}

Comprehensive modeling of hydrogen transport to the fracture process zone undoubtedly requires hydrogen trapping assessment. Of particular interest for the present study is the role played by hydrogen reversibly trapped at dislocations. Thus, the dislocation density $\rho$ is composed of the sum of the density $\rho_S$ for statistically stored dislocations (SSDs) and the density $\rho_G$ for geometrically necessary dislocations (GNDs), which are respectively associated with the macroscopic concepts of plastic strain $\varepsilon_p$ and plastic strain gradient $\varepsilon_{p,i}$. The modeling of lattice hydrogen diffusion in an iron-based material (Fig. \ref{fig:Ciron}), duplex stainless steel (Fig. \ref{fig:Cduplex}) and X-80 steel (Fig. \ref{fig:CX80}) reveals significant quantitative and qualitative differences between conventional plasticity and SGP based predictions. As GNDs do not contribute to plastic strains but to material work hardening by acting as obstacles to the motion of SSDs, incorporating their influence into the modeling leads to high levels of $c_{\ell}$ in the vicinity of the crack, where a large density of GNDs is attained to accommodate lattice curvature due to non-uniform plastic deformation. SGP predictions suggest that a critical combination of hydrogen concentration and stress will be attained very close to the crack tip, favoring hydrogen-enhanced decohesion. From a HEDE-based perspective, \citet{O09} accurately predicted crack initiation by lowering the cohesive resistance as a function of the total hydrogen concentration $c$. A linear relation between $\varepsilon_p$ and hydrogen trapped in microstructural defects was assumed in their study, leading to crack tip levels of reversibly-trapped hydrogen concentration $c_t$ one order of magnitude higher than $c_{\ell}$. Accordingly, damage nucleation (represented by collapse of the first cohesive element) occurred at the crack tip surface and not at the local stress peak (given by the conventional $\sigma_H$ distribution). Experimental measurements of high levels of surface hydrogen and critical distances of the order of micro-meters are understood, within a conventional plasticity setting, to be due to very high levels of reversibly-trapped hydrogen in the vicinity of the crack (a thorough analysis can be found in \citealp{TS01}). However, SGP-based estimations imply that the weight of $c_{\ell}$ within the total hydrogen concentration close to the crack tip could be much larger than previously anticipated and may provide some physical background to recent experimental and theoretical studies \citep{A15} that estimate a predominant role of lattice hydrogen in failure strength degradation. Fig. \ref{fig:CX80} reveals little differences between the total diffusible and the lattice hydrogen concentrations, suggesting a lesser role of reversible trapped hydrogen.\\

\begin{figure}[H]
\centering
\includegraphics[scale=0.9]{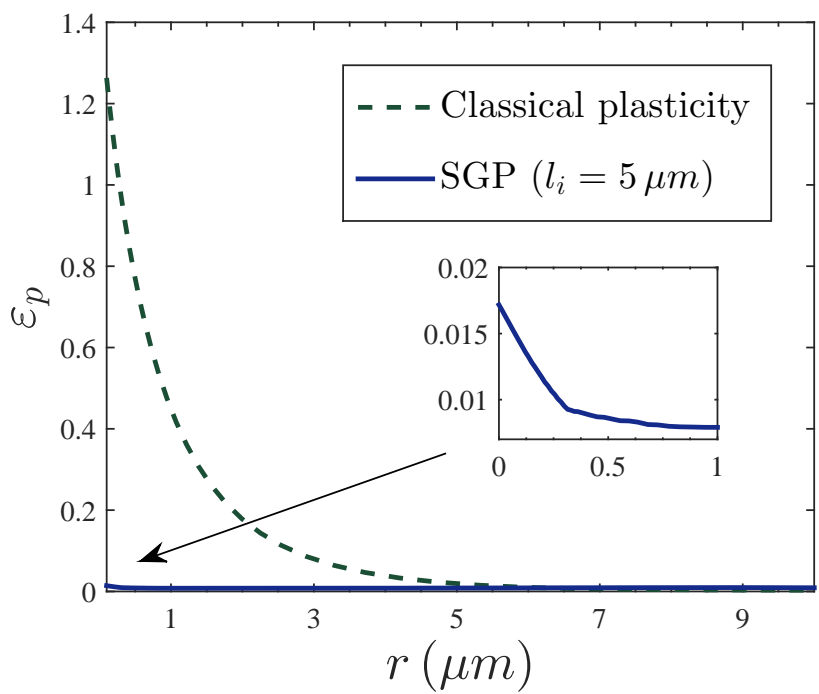}
\caption{Equivalent plastic strain distribution ahead of the crack tip in an iron-based material for an external load of $K_I= 89.7$ MPa$\sqrt{m}$ from SGP (with $l_i=5 \, \mu m$) and classical plasticity. Material properties are $E=207$ GPa, $\nu=0.3$, $\sigma_Y=250$ MPa and $N=0.2$.}
\label{fig:Ep}
\end{figure}

Physically-consistent relations between: (i) the plastic strain gradients and $\rho_G$, (ii) the plastic strains and $\rho_S$, and (iii) the total dislocation density and $c_t$, need to be established to model the kinetics of dislocation trapping accounting for both GNDs and SSDs. Large strain gradients of plastic strain close to the crack tip lead to additional hardening and lower values of $\varepsilon_p$ relative to conventional predictions. Fig. \ref{fig:Ep} shows the effective plastic strain distribution predicted by classical and gradient-enhanced plasticity models for an iron-based material under the same conditions as Fig. \ref{fig:SHiron}.

In consistency with the trends depicted by $\sigma_H$ (see Fig. \ref{fig:SHiron}), results reveal a very strong influence of GNDs within microns to the crack. This necessarily implies that gradient effects lead to a much lower SSDs density with respect to conventional plasticity predictions. However, the same argument cannot be used for $c_t$; as shown in \citep{Q04,M16}, SGP models predict large values of $\rho$ in the vicinity of the crack, as $\rho_G$ largely dominates the total dislocation density. Therefore, further research and critical experiments are need to quantitatively elucidate the role of GNDs in hydrogen trapping and other embrittlement mechanisms.

\section{Conclusions}
\label{Concluding remarks}

The role of geometrically necessary dislocations (GNDs) on crack tip hydrogen diffusion has been thoroughly investigated by means of strain gradient plasticity (SGP). The hydrostatic stress elevation and subsequent increase of hydrogen transport towards the crack tip associated with large gradients of plastic strain is examined in several metallic materials and differences with conventional plasticity quantified.\\

Results reveal a profound influence of the microstructure in several cases of particular interest from the engineering perspective. Particularly, the following key points must be highlighted:\\

- GNDs near the crack tip promote local hardening and lead to very high stresses over meaningful physical distances. The differences with classical plasticity are further enhanced in a finite strains scheme due to the contribution of strain gradients to the work hardening of the material, significantly lowering crack blunting and avoiding the local stress reduction that is observed if GNDs are neglected. A good agreement with experimental observations of crack tip deformation is observed.\\

- Very high levels of crack tip lattice hydrogen concentration are attained as a consequence of the increased dislocation density associated with gradients of plastic strain. Unlike $J_2$ plasticity-based predictions, the concentration of lattice hydrogen increases monotonically towards the crack tip.\\

- Results aim to provide insight into the embrittlement mechanisms that take place ahead of a crack. Thus, the richer description of crack tip fields provided by SGP suggests that lattice hydrogen may play a prominent role and decohesion could be readily triggered due to the high levels of stress and hydrogen concentration attained in the vicinity of the crack.

\section{Acknowledgments}
\label{Acknowledge of funding}

A. D\'{\i}az (University of Burgos) and R.P. Gangloff (University of Virginia) are acknowledged for helpful discussions. E. Mart\'{\i}nez-Pa\~neda, S. del Busto and C. Beteg\'{o}n gratefully acknowledge financial support from the Ministry of Science and Innovation of Spain through grants MAT2011-28796-CO3-03 and MAT2014-58738-C3-1-R. E. Mart\'{\i}nez-Pa\~neda also acknowledges financial support from the University of Oviedo through grant UNOV-13-PF. C. F. Niordson gratefully acknowledges financial support from the Danish Council for Independent Research under the research career programme Sapere Aude in the project ``Higher Order Theories in Solid Mechanics".




\end{document}